\documentclass[conference]{IEEEtran}
\IEEEoverridecommandlockouts
\usepackage{amsmath,amssymb,amsfonts}
\usepackage{graphicx}
\usepackage{color,colortbl}

\usepackage[export]{adjustbox}

\usepackage{placeins}
\usepackage[normalem]{ulem}

\usepackage{subfigure}

\usepackage{array}

\usepackage{multirow}
\usepackage{makecell}
\usepackage{booktabs}

\usepackage{float}

\usepackage{cite}
\usepackage{url}
\usepackage[hidelinks]{hyperref}
\hypersetup{
    colorlinks=true,
    linkcolor=black,
    filecolor=black,
    urlcolor=gray,
    citecolor=black,
    breaklinks=true
}

\usepackage{tikz}
\usepackage{pgfplots}
\pgfplotsset{width=10cm,compat=1.15}
\usepgfplotslibrary{statistics}
\usetikzlibrary{pgfplots.groupplots}

\newcommand{\etal}{\emph{et~al.}}

\newcommand{\mtrx}[1]{\mathbf{#1}}

\definecolor{mylightgray}{gray}{0.9}
\newcommand{\na}{\cellcolor{mylightgray}}

\definecolor{c1}{RGB}{0,119,187}  % Blue
\definecolor{c2}{RGB}{51,187,238} % CYAN
\definecolor{c3}{RGB}{0,153,136}  % TEAL
\definecolor{c4}{RGB}{238,119,51} % ORANGE
\definecolor{c5}{RGB}{204,51,17} % RED
\definecolor{c6}{RGB}{238,51,119} % Magenta
\definecolor{c7}{RGB}{187,187,187} % Gray
\definecolor{c8}{RGB}{0 0 0} % Black

\definecolor{ct1}{RGB}{         0,  113.9850,  188.9550}
\definecolor{ct2}{RGB}{216.7500,   82.8750,   24.9900}
\definecolor{ct3}{RGB}{0    191 230} % cyan
\definecolor{ct4}{RGB}{51   119  255}
\definecolor{ct5}{RGB}{0    0   255}    % blue
\definecolor{ct6}{RGB}{0    0   102}    % navy blue
\definecolor{ct7}{RGB}{112  112 112}    % dim gray
\definecolor{ct8}{RGB}{128  128  128} % gray
\definecolor{ct9}{RGB}{166  166  166} % dark gray 
\definecolor{ct10}{RGB}{0  153  0}  %lighet green
\definecolor{ct11}{RGB}{0  204  0}  % light green
\definecolor{ct12}{RGB}{0  255  0}  % green
\definecolor{ct13}{RGB}{255  128  0}    % orange
\definecolor{ct14}{RGB}{255  0  0}  % red

\begin{document}

\title{Audio classification of the content of food containers and drinking glasses
\thanks{This work is supported by the CHIST-ERA programme through the project CORSMAL, under UK EPSRC grant EP/S031715/1.}
}

\author{\IEEEauthorblockN{Santiago Donaher}
\IEEEauthorblockA{\textit{Centre for Intelligent Sensing} \\
\textit{Queen Mary University of London}\\
London, UK \\
s.donaher@qmul.ac.uk}
\and
\IEEEauthorblockN{Alessio Xompero}
\IEEEauthorblockA{\textit{Centre for Intelligent Sensing} \\
\textit{Queen Mary University of London}\\
London, UK \\
a.xompero@qmul.ac.uk}
\and
\IEEEauthorblockN{Andrea Cavallaro}
\IEEEauthorblockA{\textit{Centre for Intelligent Sensing} \\
\textit{Queen Mary University of London}\\
London, UK \\
a.cavallaro@qmul.ac.uk}

}

\maketitle

\begin{abstract}
Food containers, drinking glasses and cups handled by a person generate sounds that vary with the type and amount of their content. In this paper, we propose a new model for sound-based classification of the type and amount of content in a container. The proposed model is based on the decomposition of the problem into two steps, namely  action recognition and content classification. We use the scenario of the recent CORSMAL Containers Manipulation dataset and consider two actions (shaking and pouring), and seven combinations of material and filling level. The first step identifies the action performed by a person with the container. The second step determines the amount and type of content using an action-specific classifier. Experiments show that the proposed  model achieves 76.02, 78.24, and 41.89 weighted average $F_1$ score on the three test sets, respectively, and outperforms baselines and existing approaches that classify the content amount and type either independently or jointly.
\end{abstract}


\section{Introduction}
\label{sec:intro}

The contactless estimation of the weight of a food container or a drinking glass handled by a person is important to support human-robot collaborations when prior knowledge of its content is unavailable~\cite{Sanchez-Matilla2020,Medina2016,Rosenberger2021RAL}. The object weight (and hence the force to be applied by the robot hand or gripper for a human-to-robot handover) varies with the amount and type of content. We are interested in inferring this information from the sound generated by the contact of the  content with the surfaces of the container during  a pouring or a shaking action~\cite{griff,Ikeno2015,clarke,Liang2019AudioPouring}.

\begin{figure}[t!]
    \centering
    \setlength\tabcolsep{.5pt}
    \begin{tabular}{ccc}
    \includegraphics[width=.32\columnwidth]{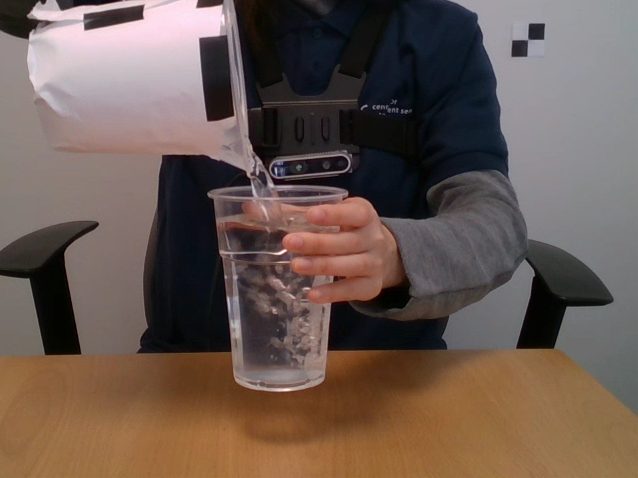} & %
    \includegraphics[width=.32\columnwidth]{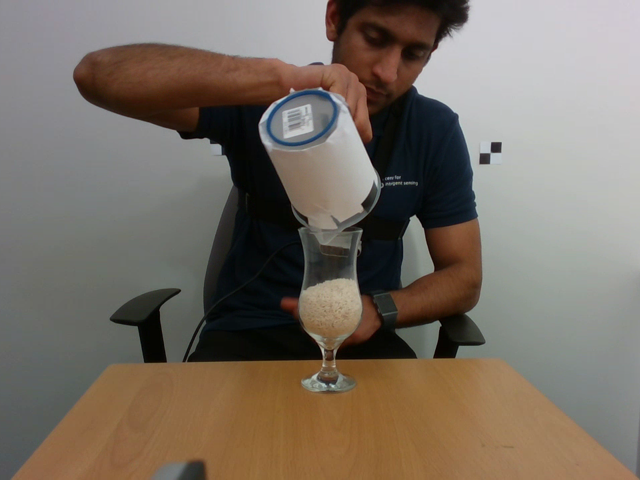} &
    \includegraphics[width=.32\columnwidth]{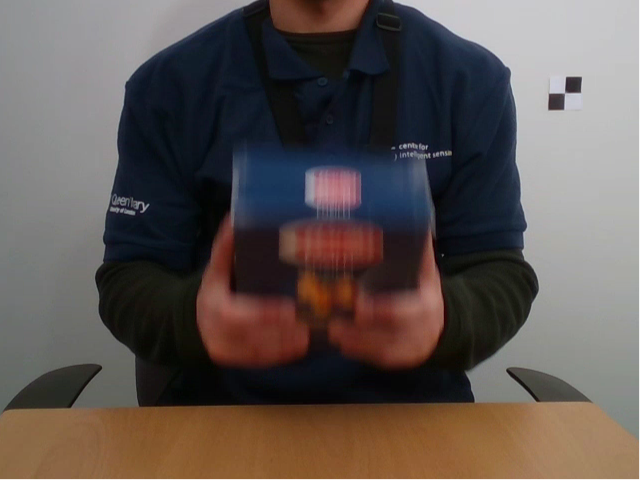} \\
    \includegraphics[height=.20\columnwidth]{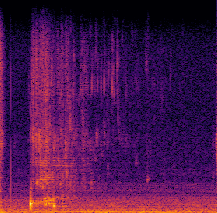} & %
    \includegraphics[height=.20\columnwidth]{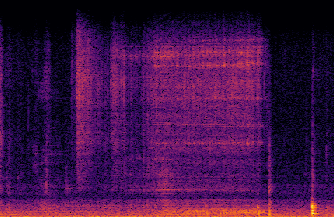} &
    \includegraphics[height=.20\columnwidth]{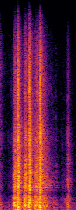} \\
    \end{tabular}
    \caption{Sample spectrograms (bottom) of the sound produced by the action of pouring  content into a drinking glass (left, centre) and shaking a food container (right).
    \vspace{-10pt}
    }
    \label{fig:spectrograms1}
\end{figure}

Previous work on audio regression of content measured the air volume variations within a container~\cite{Liang2019AudioPouring}, the vibrations of liquid poured from a bottle~\cite{Ikeno2015} and the vibrations in the acoustic frequency generated by a robot shaking and pouring granular materials from and into a tub~\cite{clarke}. 
Clarke \etal~\cite{clarke} evaluated multiple models, such as a linear regressor, a convolutional neural network, and recurrent neural networks based on either a Long Short-Term Memory (LSTM) module or a Gated Recurrent Unit (GRU) module, to regress the amount of content shaken or poured, using spectrograms as input.
Liang \etal~\cite{Liang2019AudioPouring} regressed the level of the remaining air column in a container (e.g.~glass, cup, thermo, mug) during the pouring of a liquid (water, juice) with a LSTM-based recurrent neural network. To handle noisy scenarios, Liang \etal~\cite{Liang2020MultimodalPouring} regressed the liquid level by conditioning on both audio and force/torque data of three containers, and the trained model generalises to different containers. 
However, these works specifically addressed the pouring action of containers standing on surfaces.
We are instead interested in the problem of classifying the content of a range of food containers and drinking glasses {\em handled by a person} (see Fig.~\ref{fig:spectrograms1}).

In this paper, we propose an audio classification model to recognise the properties of the content of containers manipulated by a person. We address the problem by first recognising the action occurring and then identifying an action-specific classifier to determine a set of seven combinations of content type and level. 
We validate the proposed model on the CORSMAL Containers Manipulation (CCM) dataset~\cite{Xompero_CCM}, the only publicly available dataset for the recognition of the properties of an object manipulated by a person.  In addition to CCM, we evaluate the proposed model on another set of audio recordings acquired in a different environment and with a different setup\footnote{Code, pre-trained models and data are available at: \url{http://corsmal.eecs.qmul.ac.uk/audio_classification.html}.}.

\begin{table}[t!]
\centering
\setlength\tabcolsep{.5pt}
\footnotesize
\caption{The mass of objects (container and filling) in the training split of the CCM dataset, and the capacity (cap.) of each container. The class empty corresponds to the mass of the container, which is known. key -- CX:~content type (C) and level (X), where C can be pasta (P), rice (R), or water (W); and X can be Half-full (5) or Full (9).}
\label{tab:objmasses}
\begin{tabular}{lcrrrrrrrrr}
    & &
    \includegraphics[width=.09\columnwidth]{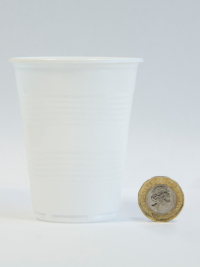} &
    \includegraphics[width=.09\columnwidth]{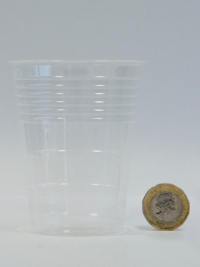} &
    \includegraphics[width=.09\columnwidth]{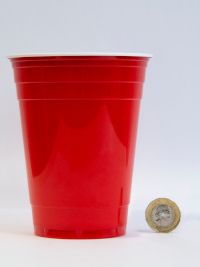} &
    \includegraphics[width=.09\columnwidth]{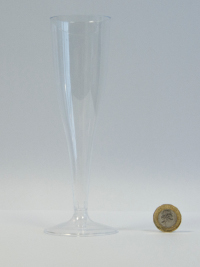} &
    \includegraphics[width=.09\columnwidth]{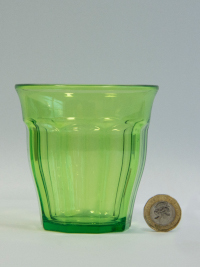} &
    \includegraphics[width=.09\columnwidth]{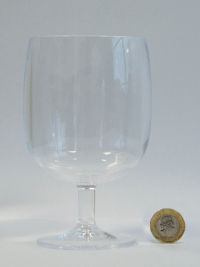} &
    \includegraphics[width=.09\columnwidth]{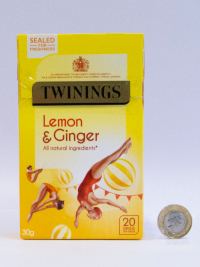} &
    \includegraphics[width=.09\columnwidth]{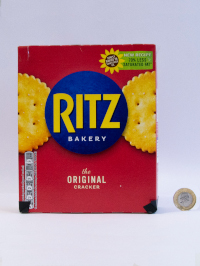} &
    \includegraphics[width=.09\columnwidth]{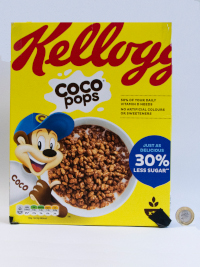} \\
    \specialrule{1.2pt}{0.2pt}{1pt} 
    \multirow{7}{*}{\parbox{0.5cm}{\textbf{Mass}\\\textbf{(g)}}}
& Empty    & 2   & 3   & 10  & 15  & 86  & 134 & 15  & 31   & 59   \\
& P5       & 37  & 39  & 119 & 37  & 131 & 205 & 104 & 297  & 776  \\
& P9       & 65  & 69  & 205 & 54  & 175 & 262 & 174 & 509  & 1350 \\
& R5       & 78  & 85  & 232 & 68  & 208 & 280 & 208 & 633  & 1548 \\
& W5       & 95  & 104 & 270 & 79  & 234 & 316 & \na & \na  & \na  \\
& R9       & 139 & 151 & 409 & 110 & 305 & 396 & 362 & 1115 & 2739 \\
& W9       & 169 & 185 & 478 & 130 & 352 & 461 & \na & \na  & \na  \\
\midrule 
\multicolumn{2}{l}{\textbf{Cap.~(mL)}}    & 185 & 202 & 520 & 128 & 296 & 363 & 472 & 1240 & 3209 \\
\specialrule{1.2pt}{0.2pt}{1pt}
\end{tabular}
\vspace{-10pt}
\end{table}

\section{Proposed method}

Let us consider a person that interacts with a container by, for example, shaking a food box or pouring content into a cup or drinking glass. During the interaction, the contact of the content with the container and other content generates different acoustic phenomena. The CORSMAL Containers Manipulation~\cite{Xompero_CCM} dataset covers this scenario with 1,140 audio recordings of 12 people manipulating 15 containers (5 cups, 5 drinking glasses, 5 food boxes) filled with {pasta}, {rice}, or {water} (note that water is not used with food boxes).

\begin{figure}[t!]
\centering
    \includegraphics[width=\columnwidth]{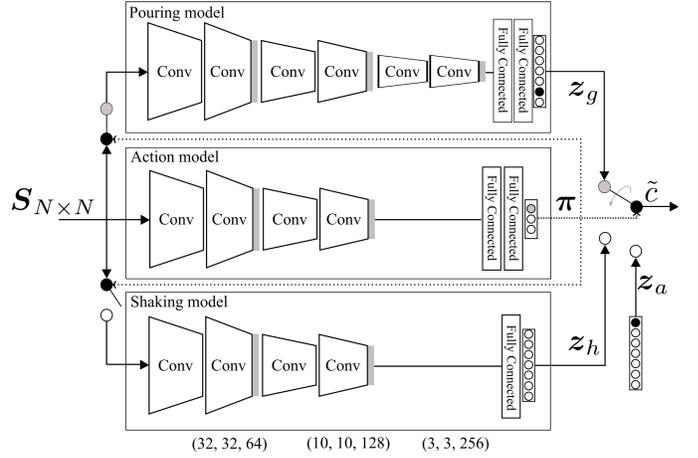}
    \caption{The proposed model for the  classification of the content type and level, $\tilde{c}$, of a container  manipulated by a person. The model first understands the action $\boldsymbol{\pi}$  based on an  $N \times N$ spectrogram $\mtrx{S}$ and then selects the appropriate branch for pouring or shaking. The size of the feature maps after each pooling layer are shown in brackets at the bottom of the diagram. Legend: {Conv:} convolutional layer,~{\protect\raisebox{2pt}{\protect\tikz \protect\draw[gray,line width=2] (0,0) -- (0.3,0);}}~{max-pooling layer}.
    }
    \label{fig:pipeline}
    % \vspace{-10pt}
\end{figure}

While each property (content type, content level) could be classified independently~\cite{Iashin2020ICPR,Ishikawa2020ICPR,Liu2020ICPR}, the combination of the two predictions can result in a wrong classification, if either is incorrect. We thus  define a set of seven classes that combine content types and levels, \mbox{$\mathcal{C} = \{ \textit{empty}, \textit{pasta-half-full}, \textit{pasta-full}, \textit{rice-half-full}, \textit{rice-full}$}, \mbox{$\textit{water-half-full}, \textit{water-full}\}$} (see Tab.~\ref{tab:objmasses}).
We tackle the task of predicting class $\tilde{c} \in \mathcal{C}$  with a model consisting of three deep neural network classifiers (see Fig.~\ref{fig:pipeline}). We named the proposed model Audio Classification of Content, or ACC. 
As pre-processing, we down-sample the signal to 22,050 \textit{Hz} and, to eliminate relative gain variations, we normalise it to [-1,1]. We perform {onset}~\cite{mcfee2015librosa} detection to identify the beginning of an action, and transform a temporal window of duration $L$ from the onset to a spectrogram $\mtrx{S}$.

The first classifier, $f(\cdot)$, is an {action model} that estimates the type of container manipulation. We also consider an \textit{unknown} class that covers cases with  no content. Thus we define the set $\mathcal{A} =\{ pouring, shaking, unknown \}$ with the classes for the action model, $f(\cdot)$, which is trained to classify the action $\boldsymbol{\pi}  \in \{0,1\}^{|\mathcal{A}|}$, where $|\cdot|$ is the cardinality of a set. The action model is a deep neural network that consists of 4 convolutional, 2 max-pooling, and 3 fully connected layers.

The second classifier is selected based on the action: $g(\cdot)$ for pouring and $h(\cdot)$ for shaking. Each specialised classifier estimates the content type and level, and is a deep neural network consisting of 6 convolutional, 3 max-pooling, and 3 fully connected layers for the pouring model; and 4 convolutional, 2 max-pooling, and 2 fully connected layers for the shaking model. For the three models, we use ReLU as activation function after each convolutional and fully connected layer, except for the last layer (softmax); max-pooling layers after every other convolutional layer; and set the kernel size for convolutional and pooling layers to 3$\times$3. The pouring model implements dropout~\cite{Srivastava2014DropoutAS} in each convolutional layer before the pooling layer to prevent overfitting.

Each specialised classifier outputs a one-hot vector $\mtrx{z} \in \{0,1\}^{|\mathcal{C}|}$, i.e.~\mbox{$\mtrx{z}_g = g(\mtrx{S})$} and \mbox{$\mtrx{z}_h = h(\mtrx{S})$}. For pouring, the classifier $g(\cdot)$ predicts the classes $\mathcal{C} \setminus \{$\textit{empty}$\}$, whereas the classifier for shaking, $h(\cdot)$ predicts the classes $\mathcal{C} \setminus \{$\textit{empty}, \textit{water-half-full}, \textit{water-full}$\}$. As the two specialised classifiers cover  situations with content only, we define an additional 7-dimensional vector $\mtrx{z}_a \in \{0,1\}^{|\mathcal{C}|}$ specifically for the \textit{unknown} cases when classifying the content properties, and that has only the first element set to 1, i.e.~\textit{empty} container for the CCM dataset.
As the decision of the action classifier conditions the choice of the classifier for the recognition of the content property (or class \textit{unknown}), we compute the final estimation as 
\begin{equation}
    \tilde{c} = \left[ \pi_1\mtrx{z}_a + \pi_2\mtrx{z}_g + \pi_3 \mtrx{z}_h \right].
\end{equation}

% We discuss the validation of ACC in the next section.

% ACC's pipeline

\section{Validation}
\label{sec:val}

\subsection{Methods under comparison}

We compare ACC with 9 alternative approaches, namely ResNet (18 layers)~\cite{He2016CVPR_ResNet}, a simpler ResNet variant (ResNet$^*$), a ResNet (18 layers) pre-trained on ImageNet (ResNet-18$^\top$)~\cite{Russakovsky2015ImageNetLS} and fine-tuned on the training split of CCM, VGG (11 layers)~\cite{vgg}, Support Vector Machine (SVM)~\cite{Cortes1995SupportN}, Random Forest~\cite{Breiman2004RandomF}, K-Nearest Neighbours (kNN)~\cite{Cover1967NearestNP}, and the top-2 submissions of the 2020 CORSMAL Challenge\footnote{\url{https://corsmal.eecs.qmul.ac.uk/challenge.html}}, namely Because It's Tactile (BIT)~\cite{Iashin2020ICPR}, and HVRL~\cite{Ishikawa2020ICPR}.
SVM, kNN, Random Forest, VGG, and ResNet-based classifiers perform direct classification as a single model.
ResNet$^*$ is a lightweight version of ResNet, consisting of 15 convolutional layers grouped into 3 blocks, with 16, 32, and 64 filters, respectively, and a fully connected layer for the 7 classes; and we also removed the first pooling layer.

BIT~\cite{Iashin2020ICPR} classifies content level and type independently with a late fusion of audio and RGB data. Audio features provided as input to a Random Forest classifier are Mel-frequency cepstrum coefficients, spectral characteristics, zero crossing rate, chroma vector and deviation; and audio features generated by a modified VGG architecture are provided as input to a GRU-based recurrent neural network. For content-type classification, BIT averages the estimation of the two classifiers. For content-level classification, BIT retrains the two classifiers and includes an additional classifier that uses spatio-temporal features extracted from the RGB frames of the four camera views in CCM. BIT then fuses the decision of the three classifiers.

HVRL~\cite{Ishikawa2020ICPR} uses a convolutional neural network based on VGG~\cite{vgg} to classify the content type from multiple sequential audio frames converted to spectrogram, followed by majority voting. For each audio frame, the feature map prior to the fully connected layers is processed by an LSTM-based recurrent neural network, followed by a fully connected layer, to temporally estimate the content level. The decision of the last frame is the final classification result.

\subsection{Datasets and performance measure}

The 1,140 audio recordings of the CCM dataset~\cite{Xompero_CCM} are split into three sets that are evenly distributed among the container types: training (684), public testing (228) and private testing (228). We randomly split the recordings of the training set into training and validation with a 80/20 ratio.  

The audio of the CCM dataset was recorded with a circular array of 8 Boya BY-M1 omnidirectional Lavelier microphones and sampling rate 44.1~kHz. The array was placed on the table and the manipulation occurred at a distance between 0.5 to 3~m from the microphones. The multi-channel sound signal is converted into mono channel by averaging samples across channels~\cite{mcfee2015librosa}.  

We also consider 21 additional recordings acquired in a different room with a Blue Yeti Studio microphone and a sampling rate of 16~bit/48~kHz. We use 4 different containers (two drinking glasses, one cup, one box) and perform a total of 19 pouring and 2 shaking actions with a different type of pasta and rice but the same filling levels as CCM. All recordings were acquired with the microphone placed on a table at a 12~cm from the container for the pouring action and about 20~cm for the shaking action.

To handle class-unbalanced data, we use the weighted average F$_1$ score, $\Bar{F}_1$, as performance measure across the $|\mathcal{C}|$ classes, each weighted by the number of audio recordings, $R_n$, in each class $n$: 
\begin{equation}
    \Bar{F}_1 = \frac{1}{R} \sum_{n=1}^{|\mathcal{C}|} R_n F_n,
    \label{eq:wafs}
\end{equation}
where $R=\sum_{n=1}^{|\mathcal{C}|} R_n$ is the total number of  recordings and $F_n$ is the F$_1$ score of class $n$, i.e.~the harmonic mean of precision (the number of true positives over the total number of true positives and false positives) and recall (the number of true positives over the total number of true positives and false negatives) for that class.
We report $\Bar{F}_1$ as a percentage for the discussion of the results.

\subsection{Implementation details}

We train each classifier independently for 100 epochs with categorical cross-entropy loss and  select the model with the minimum validation loss. During training, we use the Adam~\cite{Kingma2015ICLR_ADAM} optimiser with a learning rate of 0.001 and a batch size of 16. Since the classifiers need spectrograms of $N\times N$ resolution as input, we resize the spectrograms to $N=96$. For the temporal segmentation, we also set $L=10$~s based on the average duration of the actions. Note that we complement sound signals with duration shorter than $L$ with zero padding, as other strategies, such as repeating or interpolation, affect the prediction of the content level and worsen the accuracy in validation.

All the deep neural networks that perform direct classification are re-implemented or provided in TensorFlow. These models receive the same input as ACC. SVM, kNN, and Random Forest are provided in \textit{scikit-learn} and receive a spectrogram reshaped into a 1D vector.

\subsection{Discussion}

\begin{table}[t!]
    \centering
    % \footnotesize
    % \setlength\tabcolsep{1.5pt}
    \caption{Comparison of the weighted average $F_1$ score ($\Bar{F}_1$) of the combined content type and level classes on the test splits of ccm~\cite{Xompero_CCM} and complexity of the models. Note that bit and hvrl estimate content type and level as two independent properties. best results are in bold, second best in italic.key -- \# params.:~number of parameters.
    }
    \begin{tabular}{rlrrrr}
    \specialrule{1.2pt}{0.2pt}{1pt}
    & \textbf{Model} & \textbf{\# Params.} & \textbf{Storage} & \multicolumn{2}{c}{\textbf{CCM test sets}} \\ 
    & & $\times1000$ & MB & Public & Private \\
    \specialrule{1.2pt}{0.2pt}{1pt}
    & Random             & -- & -- & 9.57 & 11.22 \\
    \midrule
    \cite{Iashin2020ICPR}       & BIT       	       & -- & -- & 75.00 & \textit{77.86} \\
    \cite{Ishikawa2020ICPR}     & HVRL           	   & 6,839 & 82.1 & \textbf{82.14} & 73.40 \\
	\cite{Cover1967NearestNP}   & kNN               & -- & 63.5 & 70.33 & 62.99\\
    \cite{Cortes1995SupportN}   & SVM               & -- & 34.8 & 69.81 & 73.43 \\
    \cite{Breiman2004RandomF}   & Random Forest     & -- & 1.7 & 73.59 & 70.48 \\
    \cite{vgg}                  & VGG-11            & 44,907 & 175.4 & 74.03 & 72.67 \\
    \cite{He2016CVPR_ResNet}    & ResNet$^*$ & 179 & 0.8 & 70.45 & 71.56 \\
    \cite{He2016CVPR_ResNet}    & ResNet-18          & 11,692 & 45.8 & 62.28 & 55.28 \\
    \cite{He2016CVPR_ResNet}    & ResNet-18$^\top$   & 11,692 & 45.8 & 59.69 & 63.01 \\
    \midrule
    & \textbf{ACC}     & 16,482 & 64.3 & \textit{76.02} & \textbf{78.24} \\
    \specialrule{1.2pt}{0.2pt}{1pt}
    \end{tabular}
    \vspace{-10pt}
    \label{tab:accplotjointmain}
\end{table}
%==================================================

Tab.~\ref{tab:accplotjointmain} compares the weighted average $F_1$ scores of the models under analysis on the CCM test splits, and their complexity in terms of number of parameters and storage size. ACC achieves $\Bar{F}_1$ of 76.02 and 78.24 for the two test splits, respectively. ACC outperforms all the models that directly classify on both test splits. 
HVRL is the best performing on the public test split, but its accuracy decrease on the private test split shows that independently estimating the two content properties does not always guarantee consistency of the two predictions. BIT classifies content types and levels independently, and shows a similar behaviour of HVRL but achieving higher weighted average $F_1$ score in the private test split and a lower one in the public test split.

The architecture of ACC is based on that of VGG but with almost 3 times fewer parameters (16,482,637 versus 44,907,000, respectively), even though three independent classifiers are used: the action classifier has 7,077,315 parameters, the classifier specific for pouring has 2,590,150 parameters, and the classifier specific for shaking has 6,815,172 parameters. Moreover, ACC has 5,000,000 more parameters than the two ResNet-18 models, whereas HVRL has a convolutional neural network of  4,472,580 and an LSTM of 2,366,211 parameters, respectively, that requires more storage (82.1 MB). However, the modified ResNet has only 179,000 parameters, less than 1 MB in storage, and achieves higher performance than ResNet-18. Note also how the use of the resized spectrogram as input during training makes the storage of the trained kNN and SVM classifiers comparable to the deep neural network classifiers, while Random Forest better copes with the large input and needs only 1.7 MB to store the trained model.

\pgfplotstableread{fig4_01.txt}\confmatknn
\pgfplotstableread{fig4_02.txt}\confmatsvm
\pgfplotstableread{fig4_03.txt}\confmatrf
\pgfplotstableread{fig4_04.txt}\confmatrna
\pgfplotstableread{fig4_05.txt}\confmatrnb
\pgfplotstableread{fig4_06.txt}\confmatrnbtl
\pgfplotstableread{fig4_07.txt}\confmatvgg
\pgfplotstableread{fig4_08.txt}\confmatscc

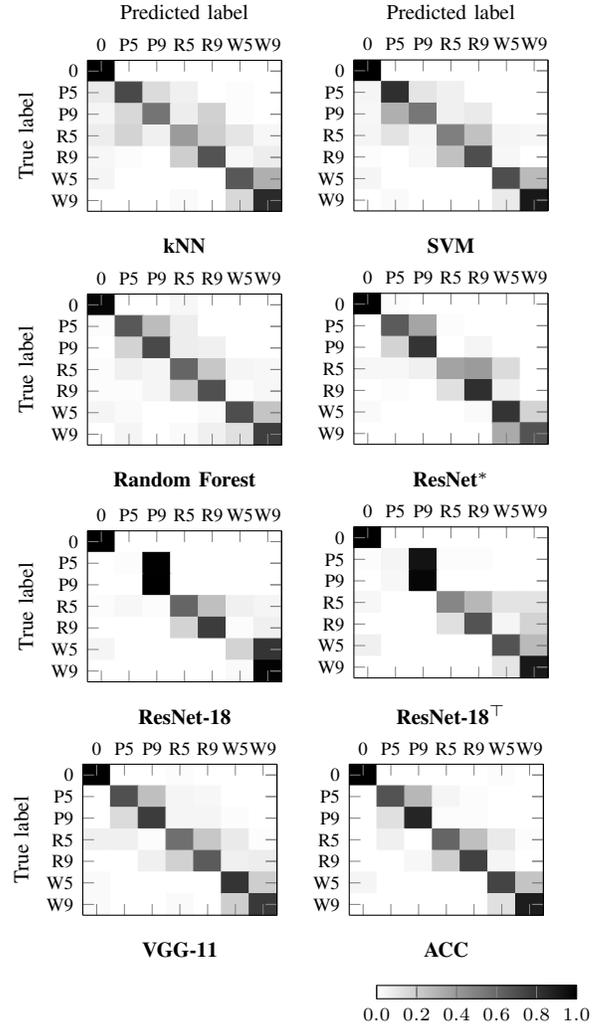
\begin{figure}[t!]
    \centering
    \begin{tikzpicture}
        \begin{axis}[
            enlargelimits=false,
            axis on top,
            axis x line*=bottom,
            width=.47\columnwidth,
            ylabel=True label,
            xlabel=\textbf{kNN},
            view={0}{90},
            y dir=reverse,
            point meta min=0,
            point meta max=1,
            xmin=-0.5,xmax=6.5,
            ymin=-0.5,ymax=6.5,
            ytick={0,1,2,3,4,5,6},
            xtick={0,1,2,3,4,5,6},
            xticklabels={},
            yticklabels={0,P5,P9,R5,R9,W5,W9},
            tick label style={font=\scriptsize},
            label style={font=\footnotesize},
            colormap={bw}{gray(0cm)=(1);gray(1cm)=(0);},
        ]
        \addplot [matrix plot*,mesh/cols=7, point meta=explicit] table [meta=z] {\confmatknn};
        \end{axis}
        \begin{axis}[
            axis x line*=top,
            axis y line*=right,
            y dir=reverse,
            width=.47\columnwidth,
            view={0}{90},   %
            xlabel=Predicted label,
            enlargelimits=false,
            axis on top,
            point meta min=0,
            point meta max=1,
            xmin=-0.5,xmax=6.5,
            ymin=-0.5,ymax=6.5,
            ytick={0,1,2,3,4,5,6},
            xtick={0,1,2,3,4,5,6},
            xticklabels={0,P5,P9,R5,R9,W5,W9},
            yticklabels={},
            tick label style={font=\scriptsize},
            yticklabel style = {},
            label style={font=\footnotesize},
         ]
         \end{axis}
    \end{tikzpicture}
    \begin{tikzpicture}
        \begin{axis}[
            enlargelimits=false,
            axis on top,
            axis x line*=bottom,
            width=.47\columnwidth,
            xlabel=\textbf{SVM},
            view={0}{90},
            y dir=reverse,
            point meta min=0,
            point meta max=1,
            xmin=-0.5,xmax=6.5,
            ymin=-0.5,ymax=6.5,
            ytick={0,1,2,3,4,5,6},
            xtick={0,1,2,3,4,5,6},
            xticklabels={},
            yticklabels={0,P5,P9,R5,R9,W5,W9},
            tick label style={font=\scriptsize},
            label style={font=\footnotesize},
            colormap={bw}{gray(0cm)=(1);gray(1cm)=(0);},
        ]
        \addplot [matrix plot*,mesh/cols=7, point meta=explicit] table [meta=z] {\confmatsvm};
        \end{axis}
        \begin{axis}[
            axis x line*=top,
            axis y line*=right,
            y dir=reverse,
            width=.47\columnwidth,
            view={0}{90},   %
            xlabel=Predicted label,
            enlargelimits=false,
            axis on top,
            point meta min=0,
            point meta max=1,
            xmin=-0.5,xmax=6.5,
            ymin=-0.5,ymax=6.5,
            ytick={0,1,2,3,4,5,6},
            xtick={0,1,2,3,4,5,6},
            xticklabels={0,P5,P9,R5,R9,W5,W9},
            yticklabels={},
            tick label style={font=\scriptsize},
            yticklabel style = {},
            label style={font=\footnotesize},
         ]
         \end{axis}
    \end{tikzpicture}
    \begin{tikzpicture}
        \begin{axis}[
            enlargelimits=false,
            axis on top,
            axis x line*=bottom,
            width=.47\columnwidth,
            ylabel=True label,
            xlabel=\textbf{Random Forest},
            view={0}{90},
            y dir=reverse,
            point meta min=0,
            point meta max=1,
            xmin=-0.5,xmax=6.5,
            ymin=-0.5,ymax=6.5,
            ytick={0,1,2,3,4,5,6},
            xtick={0,1,2,3,4,5,6},
            xticklabels={},
            yticklabels={0,P5,P9,R5,R9,W5,W9},
            tick label style={font=\scriptsize},
            label style={font=\footnotesize},
            colormap={bw}{gray(0cm)=(1);gray(1cm)=(0);},
        ]
        \addplot [matrix plot*,mesh/cols=7, point meta=explicit] table [meta=z] {\confmatrf};
        \end{axis}
        \begin{axis}[
            axis x line*=top,
            axis y line*=right,
            y dir=reverse,
            width=.47\columnwidth,
            view={0}{90},   %
            enlargelimits=false,
            axis on top,
            point meta min=0,
            point meta max=1,
            xmin=-0.5,xmax=6.5,
            ymin=-0.5,ymax=6.5,
            ytick={0,1,2,3,4,5,6},
            xtick={0,1,2,3,4,5,6},
            xticklabels={0,P5,P9,R5,R9,W5,W9},
            yticklabels={},
            tick label style={font=\scriptsize},
            yticklabel style = {},
            label style={font=\footnotesize},
         ]
         \end{axis}
    \end{tikzpicture}
    \begin{tikzpicture}
        \begin{axis}[
            enlargelimits=false,
            axis on top,
            axis x line*=bottom,
            width=.47\columnwidth,
            xlabel=\textbf{ResNet$^*$},
            view={0}{90},
            y dir=reverse,
            point meta min=0,
            point meta max=1,
            xmin=-0.5,xmax=6.5,
            ymin=-0.5,ymax=6.5,
            ytick={0,1,2,3,4,5,6},
            xtick={0,1,2,3,4,5,6},
            xticklabels={},
            yticklabels={0,P5,P9,R5,R9,W5,W9},
            tick label style={font=\scriptsize},
            label style={font=\footnotesize},
            colormap={bw}{gray(0cm)=(1);gray(1cm)=(0);},
        ]
        \addplot [matrix plot*,mesh/cols=7, point meta=explicit] table [meta=z] {\confmatrna};
        \end{axis}
        \begin{axis}[
            axis x line*=top,
            axis y line*=right,
            y dir=reverse,
            width=.47\columnwidth,
            view={0}{90},   %
            enlargelimits=false,
            axis on top,
            point meta min=0,
            point meta max=1,
            xmin=-0.5,xmax=6.5,
            ymin=-0.5,ymax=6.5,
            ytick={0,1,2,3,4,5,6},
            xtick={0,1,2,3,4,5,6},
            xticklabels={0,P5,P9,R5,R9,W5,W9},
            yticklabels={},
            tick label style={font=\scriptsize},
            yticklabel style = {},
            label style={font=\footnotesize},
         ]
         \end{axis}
    \end{tikzpicture}
    \begin{tikzpicture}
        \begin{axis}[
            enlargelimits=false,
            axis on top,
            axis x line*=bottom,
            width=.47\columnwidth,
            ylabel=True label,
            xlabel=\textbf{ResNet-18},
            view={0}{90},
            y dir=reverse,
            point meta min=0,
            point meta max=1,
            xmin=-0.5,xmax=6.5,
            ymin=-0.5,ymax=6.5,
            ytick={0,1,2,3,4,5,6},
            xtick={0,1,2,3,4,5,6},
            xticklabels={},
            yticklabels={0,P5,P9,R5,R9,W5,W9},
            tick label style={font=\scriptsize},
            label style={font=\footnotesize},
            colormap={bw}{gray(0cm)=(1);gray(1cm)=(0);},
        ]
        \addplot [matrix plot*,mesh/cols=7, point meta=explicit] table [meta=z] {\confmatrnb};
        \end{axis}
        \begin{axis}[
            axis x line*=top,
            axis y line*=right,
            y dir=reverse,
            width=.47\columnwidth,
            view={0}{90},   %
            enlargelimits=false,
            axis on top,
            point meta min=0,
            point meta max=1,
            xmin=-0.5,xmax=6.5,
            ymin=-0.5,ymax=6.5,
            ytick={0,1,2,3,4,5,6},
            xtick={0,1,2,3,4,5,6},
            xticklabels={0,P5,P9,R5,R9,W5,W9},
            yticklabels={},
            tick label style={font=\scriptsize},
            yticklabel style = {},
            label style={font=\footnotesize},
         ]
         \end{axis}
    \end{tikzpicture}
    \begin{tikzpicture}
        \begin{axis}[
            enlargelimits=false,
            axis on top,
            axis x line*=bottom,
            width=.47\columnwidth,
            xlabel=\textbf{ResNet-18$^\top$},
            view={0}{90},
            y dir=reverse,
            point meta min=0,
            point meta max=1,
            xmin=-0.5,xmax=6.5,
            ymin=-0.5,ymax=6.5,
            ytick={0,1,2,3,4,5,6},
            xtick={0,1,2,3,4,5,6},
            xticklabels={},
            yticklabels={0,P5,P9,R5,R9,W5,W9},
            tick label style={font=\scriptsize},
            label style={font=\footnotesize},
            colormap={bw}{gray(0cm)=(1);gray(1cm)=(0);},
        ]
        \addplot [matrix plot*,mesh/cols=7, point meta=explicit] table [meta=z] {\confmatrnbtl};
        \end{axis}
        \begin{axis}[
            axis x line*=top,
            axis y line*=right,
            y dir=reverse,
            width=.47\columnwidth,
            view={0}{90},   %
            enlargelimits=false,
            axis on top,
            point meta min=0,
            point meta max=1,
            xmin=-0.5,xmax=6.5,
            ymin=-0.5,ymax=6.5,
            ytick={0,1,2,3,4,5,6},
            xtick={0,1,2,3,4,5,6},
            xticklabels={0,P5,P9,R5,R9,W5,W9},
            yticklabels={},
            tick label style={font=\scriptsize},
            yticklabel style = {},
            label style={font=\footnotesize},
         ]
         \end{axis}
    \end{tikzpicture}
    \begin{tikzpicture}
        \begin{axis}[
            enlargelimits=false,
            axis on top,
            axis x line*=bottom,
            width=.47\columnwidth,
            ylabel=True label,
            xlabel=\textbf{VGG-11},
            view={0}{90},
            y dir=reverse,
            point meta min=0,
            point meta max=1,
            xmin=-0.5,xmax=6.5,
            ymin=-0.5,ymax=6.5,
            ytick={0,1,2,3,4,5,6},
            xtick={0,1,2,3,4,5,6},
            xticklabels={},
            yticklabels={0,P5,P9,R5,R9,W5,W9},
            tick label style={font=\scriptsize},
            label style={font=\footnotesize},
            colormap={bw}{gray(0cm)=(1);gray(1cm)=(0);},
        ]
        \addplot [matrix plot*,mesh/cols=7, point meta=explicit] table [meta=z] {\confmatvgg};
        \end{axis}
        \begin{axis}[
            axis x line*=top,
            axis y line*=right,
            y dir=reverse,
            width=.47\columnwidth,
            view={0}{90},   %
            enlargelimits=false,
            axis on top,
            point meta min=0,
            point meta max=1,
            xmin=-0.5,xmax=6.5,
            ymin=-0.5,ymax=6.5,
            ytick={0,1,2,3,4,5,6},
            xtick={0,1,2,3,4,5,6},
            xticklabels={0,P5,P9,R5,R9,W5,W9},
            yticklabels={},
            tick label style={font=\scriptsize},
            yticklabel style = {},
            label style={font=\footnotesize},
         ]
         \end{axis}
    \end{tikzpicture}
    \begin{tikzpicture}
        \begin{axis}[
            enlargelimits=false,
            axis on top,
            axis x line*=bottom,
            width=.47\columnwidth,
            xlabel=\textbf{ACC},
            view={0}{90},
            y dir=reverse,
            point meta min=0,
            point meta max=1,
            xmin=-0.5,xmax=6.5,
            ymin=-0.5,ymax=6.5,
            ytick={0,1,2,3,4,5,6},
            xtick={0,1,2,3,4,5,6},
            xticklabels={},
            yticklabels={0,P5,P9,R5,R9,W5,W9},
            tick label style={font=\scriptsize},
            label style={font=\footnotesize},
            colormap={bw}{gray(0cm)=(1);gray(1cm)=(0);},
        ]
        \addplot [matrix plot*,mesh/cols=7, point meta=explicit] table [meta=z] {\confmatscc};
        \end{axis}
        \begin{axis}[
            axis x line*=top,
            axis y line*=right,
            y dir=reverse,
            width=.47\columnwidth,
            view={0}{90},   %
            enlargelimits=false,
            axis on top,
            point meta min=0,
            point meta max=1,
            xmin=-0.5,xmax=6.5,
            ymin=-0.5,ymax=6.5,
            ytick={0,1,2,3,4,5,6},
            xtick={0,1,2,3,4,5,6},
            xticklabels={0,P5,P9,R5,R9,W5,W9},
            yticklabels={},
            tick label style={font=\scriptsize},
            yticklabel style = {},
            label style={font=\footnotesize},
         ]
         \end{axis}
    \end{tikzpicture}
    \vspace{7pt}
    \\
    \begin{tikzpicture}
    \begin{axis}[
            hide axis,  
            scale only axis,
            height=.05\columnwidth, width=0.3\columnwidth,
            anchor=south west,
            point meta min=0,
            point meta max=1,
            colorbar horizontal,
            colorbar style={
                height=.02\columnwidth, 
                at={(2.5,1.05)},
                anchor=south west,
                xticklabel pos=lower,
                xticklabel style={
                    /pgf/number format/.cd,
                    fixed,
                    precision=1,
                    fixed zerofill,
                    tick label style={font=\scriptsize},
                },
            },
            colormap={bw}{gray(0cm)=(1);gray(1cm)=(0);},
        ]
        \addplot [draw=none] coordinates {(0,0)};
        \end{axis}
    \end{tikzpicture}
    \caption{Confusion matrices for all methods across all the containers of the public and private testing splits of CCM~\cite{Xompero_CCM}. KEY -- 0:~empty; CX:~content type (C) and level (X), where C represents pasta (P), rice (R), or water (W); and X is  half-full (5) or full (9).
    \vspace{-10pt}
    }
    \label{fig:confmatpaper}
\end{figure}

Fig.~\ref{fig:confmatpaper} compares the confusion matrices for the models performing classification of the combined content type and level on the combined CCM test splits. ResNet-18 and ResNet-18$^\top$ wrongly classifies \textit{pasta-half-full} with \textit{pasta-full}, and transfer learning helps to reduce the wrong classification of \textit{water-half-full} with \textit{water-full}. Reducing the number of layers from 18 to 15 helps to increase the correct classification across the different classes, while decreasing the overfitting on this small-size dataset. Overall, the methods tend to wrongly classify the level \textit{half-full} with \textit{full} and the opposite, while the \textit{empty} (\textit{unknown}) case is correctly classified by all methods. ACC reduces the wrong classifications and achieves a higher number of correct classifications along the diagonal compared to the other models.

\begin{table}[t!]
\centering
\footnotesize
\caption{Comparison of the Weighted Average $F_1$ score ($\Bar{F}_1$) on the testing set with 21 audio recordings of novel containers used in a different environment and setup. Best results are in bold, second best in italic.
% \vspace{-5pt}
}
\begin{tabular}{rl ccc}
\specialrule{1.2pt}{0.2pt}{1pt}
\multicolumn{2}{c}{\textbf{Model}} & \textbf{Level}& \textbf{Type} & \textbf{Type \& Level}\\ 
\specialrule{1.2pt}{0.2pt}{1pt}
                         	 & Random       	& 43.98	& 17.05	 &  15.24 \\
\midrule				
\cite{Cover1967NearestNP}	 & kNN          	& 37.28	& 37.06	 & 19.09 \\
\cite{Cortes1995SupportN}	 & SVM          	& 40.77	 & 47.39    	 & 26.98 \\
\cite{Breiman2004RandomF}	 & Random Forest    & 47.62	& 44.49	 & 20.07 \\
\cite{vgg}	 & VGG-11             	& \textbf{58.53}	& 25.95	 & 15.24 \\
\cite{He2016CVPR_ResNet} 	 & ResNet$^*$          	& 52.04 &	\textit{56.11}	& \textit{34.95} \\
\cite{He2016CVPR_ResNet} 	 & ResNet-18          	& 50.40	& 45.31	 & 30.87 \\
\cite{He2016CVPR_ResNet} 	 & ResNet-18$^\top$      	& 50.30	& \textbf{61.50}	 & 32.02 \\
\midrule				
                         	 & \textbf{ACC}          	 & \textit{58.34}     	& 53.85	 & \textbf{41.89} \\

\specialrule{1.2pt}{0.2pt}{1pt}
\end{tabular}
\label{tab:comparisonsnew}
% \vspace{-10pt}
\end{table}

Tab.~\ref{tab:comparisonsnew} compares the results obtained from the recordings of the second, additional setup. ACC achieves $\Bar{F}_1=58.34$ in correctly predicting only the content level, outperforming the other approaches except VGG-11 ($\Bar{F}_1=58.53$). For content type, ACC achieves $\Bar{F}_1=53.85$ and only ResNet-18 and ResNet$^*$ achieves higher accuracy ($\Bar{F}_1=61.50$ and $56.11$, respectively). Despite this difference between the accuracy for content type and content level, ACC outperforms other approaches when validating the combination of both content type and level. ACC achieves $\Bar{F}_1=41.89$, while the second best, ResNet$^*$, achieves $\Bar{F}_1=34.95$.

Fig.~\ref{fig:confmat_action} shows the confusion matrices of the action classifier for the combined test sets in CCM and the test set acquired in a different environment and setup. The action classifier of ACC can correctly distinguish between the two actions of pouring and shaking, as well as the \textit{unknown} case, for unknown containers in the same setup and environment (CCM). The prediction is instead more challenging in the novel setup and environment, as \textit{unknown} is confused with \textit{pouring} and some recordings with \textit{pouring} are confused with \textit{shaking}. 
The higher sensitivity and closer position of the Blue Yeti Studio microphone compared to the Boya BY-M1 omnidirectional Lavelier microphone are possible causes of the lower performance in addition to the type of content.

\pgfplotstableread{fig5_01.txt}\actionprivtest
\pgfplotstableread{fig5_02.txt}\actionnovelcont
\begin{figure}[t!]
    \centering
        \begin{tikzpicture}
        \begin{axis}[
            enlargelimits=false,
            axis on top,
            axis x line*=bottom,
            width=.5\columnwidth,
            ylabel=True label,
            xlabel=\textbf{CCM},
            view={0}{90},
            y dir=reverse,
            point meta min=0,
            point meta max=1,
            xmin=-0.5,xmax=2.5,
            ymin=-0.5,ymax=2.5,
            ytick={0,1,2},
            xtick={0,1,2},
            xticklabels={},
            yticklabels={U,P,S},
            tick label style={font=\scriptsize},
            label style={font=\footnotesize},
            colormap={bw}{gray(0cm)=(1);gray(1cm)=(0);},
        ]
        \addplot [matrix plot*,mesh/cols=3, point meta=explicit] table [meta=z] {\actionprivtest};
        \end{axis}
        \begin{axis}[
            axis x line*=top,
            axis y line*=right,
            y dir=reverse,
            width=.5\columnwidth,
            view={0}{90},   %
            xlabel=Predicted label,
            enlargelimits=false,
            axis on top,
            point meta min=0,
            point meta max=1,
            xmin=-0.5,xmax=2.5,
            ymin=-0.5,ymax=2.5,
            ytick={0,1,2},
            xtick={0,1,2},
            xticklabels={U,P,S},
            yticklabels={},
            tick label style={font=\scriptsize},
            yticklabel style = {},
            label style={font=\footnotesize},
         ]
         \end{axis}
    \end{tikzpicture}
    \begin{tikzpicture}
        \begin{axis}[
            enlargelimits=false,
            axis on top,
            axis x line*=bottom,
            width=.5\columnwidth,
            ylabel=True label,
            xlabel=\textbf{NC},
            view={0}{90},
            y dir=reverse,
            xmin=-0.5,xmax=2.5,
            ymin=-0.5,ymax=2.5,
            ytick={0,1,2},
            xtick={0,1,2},
            xticklabels={},
            yticklabels={U,P,S},
            tick label style={font=\scriptsize},
            label style={font=\footnotesize},
            colormap={bw}{gray(0cm)=(1);gray(1cm)=(0);},
        ]
        \addplot [matrix plot*,mesh/cols=3, point meta=explicit] table [meta=z] {\actionnovelcont};
        \end{axis}
        \begin{axis}[
            axis x line*=top,
            axis y line*=right,
            y dir=reverse,
            width=.5\columnwidth,
            view={0}{90},   %
            xlabel=Predicted label,
            enlargelimits=false,
            axis on top,
            point meta min=0,
            point meta max=1,
            xmin=-0.5,xmax=2.5,
            ymin=-0.5,ymax=2.5,
            ytick={0,1,2},
            xtick={0,1,2},
            xticklabels={U,P,S},
            yticklabels={},
            tick label style={font=\scriptsize},
            yticklabel style = {},
            label style={font=\footnotesize},
         ]
         \end{axis}
    \end{tikzpicture}
    \vspace{7pt}
    \\
    \begin{tikzpicture}
    \begin{axis}[
            hide axis,  
            scale only axis,
            height=.05\columnwidth, width=0.3\columnwidth,
            anchor=south west,
            point meta min=0,
            point meta max=1,
            colorbar horizontal,
            colorbar style={
                height=.02\columnwidth, 
                at={(2.5,1.05)},
                anchor=south west,
                xticklabel pos=lower,
                xticklabel style={
                    /pgf/number format/.cd,
                    fixed,
                    precision=1,
                    fixed zerofill,
                    tick label style={font=\scriptsize},
                },
            },
            colormap={bw}{gray(0cm)=(1);gray(1cm)=(0);},
        ]
        \addplot [draw=none] coordinates {(0,0)};
        \end{axis}
    \end{tikzpicture}
    \caption{Confusion matrices for the action classifier of ACC across all the containers of the testing split in CCM (left) and the novel containers (NC) used in the second setup (right). KEY -- U:~unknown, P:~pouring, S:~shaking. 
    \vspace{-10pt}
    }
    \label{fig:confmat_action}
\end{figure}
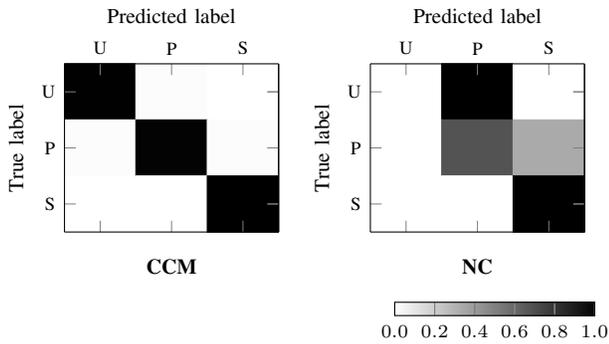

%%%%%%%%%%%%%%%%%%%%%%%%%%%
\section{Conclusion}
\label{sec:concl}

We presented an audio classification model that recognises the content type and level in a container that  is manipulated by a person who shakes it or  pours content into it. The proposed model works with both liquid and granular materials and classifies the  properties of a range of unknown containers with different materials, shapes and sizes. The proposed model outperforms existing solutions for the estimation of content type and level independently, as well as baselines based on SVM, kNN, random forest and deep neural networks that directly classify content type and level. As future work, we will extend the proposed model to generalise to different setups and environments, and to cope with multi-modal data~\cite{Liang2019AudioPouring}.

% References should be produced using the bibtex program from suitable
% BiBTeX files (here: strings, refs, manuals). The IEEEbib.bst bibliography
% style file from IEEE produces unsorted bibliography list.
% -------------------------------------------------------------------------
\bibliographystyle{IEEEbib}
\bibliography{strings,refs}

% -------------------------------------------------------------------------

\end{document}